\documentclass[letterpaper, 10 pt, conference]{ieeeconf}  

\IEEEoverridecommandlockouts                              

\usepackage{cite}
\usepackage{amsmath,amssymb,amsfonts}
\usepackage{algorithmic}
\usepackage{graphicx}
\usepackage{textcomp}
\usepackage{xcolor}
\usepackage[hidelinks]{hyperref}
\usepackage{subcaption}
\usepackage{booktabs}
\usepackage{multirow}

\usepackage{orcidlink}
\usepackage{pifont}

\title{\LARGE \bf
Evaluating Cross-Subject and Cross-Device Consistency \\in Visual Fixation Prediction
}

\author{Yuli Wu$\,$\orcidlink{0000-0002-6216-4911}$^{\,1\,}$, Henning Konermann$\;$\orcidlink{0009-0006-5736-5803}$^{\,1\,}$, Emil Mededovic$\,$\orcidlink{0009-0002-6801-4890}$^{\,1\,}$, Peter Walter$\,$\orcidlink{0000-0001-8745-6593}$^{\,2\,}$ and Johannes Stegmaier$\,$\orcidlink{0000-0003-4072-3759}$^{\,1\,}$
\thanks{This work was supported by Deutsche Forschungsgemeinschaft (DFG, German Research Foundation) with the grants GRK2610: InnoRetVision (HK, YW, project number 424556709) and FOR2591: Severity Assessment in Animal-Based Research (EM, project number 321137804).}%
\thanks{$^{1}\,$YW, HK, EM and JS are with the Institute of Imaging and Computer Vision, RWTH Aachen University, Aachen, Germany. Correspondence: {\tt\small yuli.wu@lfb.rwth-aachen.de } }%
\thanks{$^{2}\,$PW is with the Department of Ophthalmology, RWTH Aachen University, Aachen, Germany.}%
}

\begin{document}

\maketitle
\thispagestyle{empty}
\pagestyle{empty}

\begin{abstract}
Understanding cross-subject and cross-device consistency in visual fixation prediction is essential for advancing eye-tracking applications, including visual attention modeling and neuroprosthetics. This study evaluates fixation consistency using an embedded eye tracker integrated into regular-sized glasses, comparing its performance with high-end standalone eye-tracking systems. Nine participants viewed 300 images from the MIT1003 dataset in subjective experiments, allowing us to analyze cross-device and cross-subject variations in fixation patterns with various evaluation metrics. Our findings indicate that average visual fixations can be reliably transferred across devices for relatively simple stimuli. However, individual-to-average consistency remains weak, highlighting the challenges of predicting individual fixations across devices. These results provide an empirical foundation for leveraging predicted average visual fixation data to enhance neuroprosthetic applications.
\newline

\indent \textit{Clinical relevance}— Cross-device and cross-subject consistency in visual fixations can improve visual prosthetics for individuals with visual impairments by enabling reliable eye-gaze-assisted functionality.
\end{abstract}


\section{INTRODUCTION}
\begin{figure}[b!]
    \centering
    \includegraphics[width=0.85\linewidth]{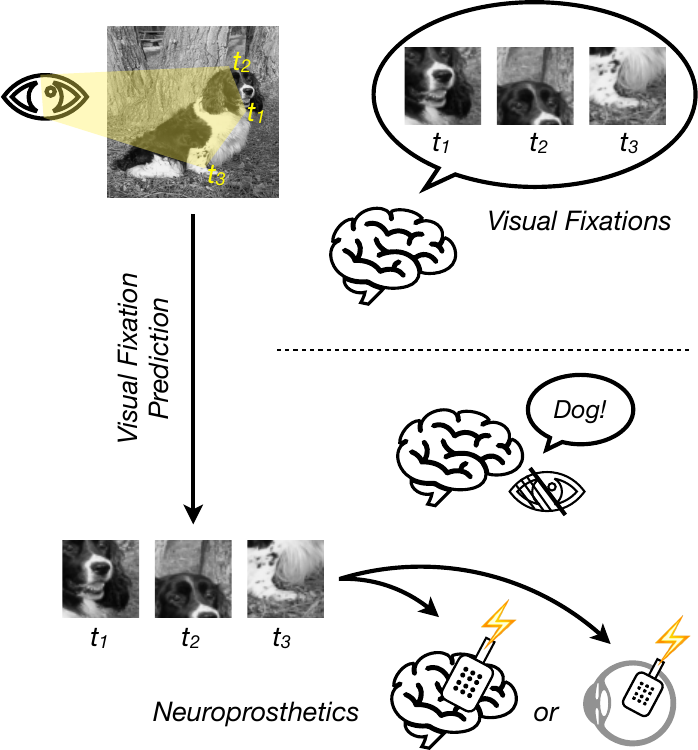}\vspace{0.2em}
    \caption{Application of the visual fixation prediction on neuroprosthetics.}
    \label{fig:motivation}
\end{figure}

Visual fixation patterns \cite{yarbus1967eye} serve as key indicators of attentional processes in visual tasks and are widely utilized in cognitive research \cite{land1999roles,li_saliency_2002,torralba2006contextual,golan2020controversial}. Beyond cognition, they find applications in medical image analysis \cite{ma2024eyegaze}, autism identification \cite{jiang2017learning}, Parkinson’s disease diagnosis \cite{koch2024eye}, and visual design assessment \cite{urano2021visual}. Recent studies leverage visual fixation information to enhance the precision of neuroprosthetics \cite{wu_fixational_2024,wu2025visual}, which can be integrated into existing neural encoding approaches for visual prosthetics, e.g., \cite{wu2023deep,wu2024optimizing}. Achieving this requires predicting average visual fixations from a group of subjects, as individuals with visual impairments may lack eye movements, as illustrated in Fig. \ref{fig:motivation}.

However, variability in device specifications, calibration methods, and sampling rates across different eye-tracking systems introduces inconsistencies in gaze data. Additionally, individual behavioral differences challenge the generalization of fixation patterns. This study hypothesizes that visual fixation patterns exhibit cross-device and cross-subject consistency when processed with standardized methods, as core visual attention mechanisms should lead to stable fixation trends regardless of device or participant variability.

Deep learning-based gaze prediction, particularly the DeepGaze family, utilizes CNNs and Transformers to estimate gaze from eye images. Early models like DeepGaze I \cite{DeepGazeI} relied on CNNs, while DeepGaze II \cite{DeepGazeII} incorporated head pose and saliency maps. DeepGaze IIe \cite{DeepGazeIIE} further enhanced performance with eye-region attention mechanisms. More recent models, such as DeepGaze III \cite{DeepGazeIII}, leverage Transformers for improved generalization, achieving state-of-the-art performance without specialized hardware. Additionally, the self-attention maps in Transformer architectures \cite{NIPS2017_3f5ee243} can be interpreted as saliency maps, enabling the adaptation of pre-trained foundation models like DINO \cite{caron2021emerging} and DINOv2 \cite{oquabdinov2} for saliency prediction. 

In this study, we conduct subjective experiments using eye trackers embedded in regular-sized glasses to evaluate visual fixation consistency in comparison to high-end standalone eye trackers. Additionally, we assess cross-subject consistency with this wearable eye-tracking device, laying the foundation for potential applications based on subjective assessments. To support further research, we have publicly released visual fixation data from 9 subjects on 300 images at \url{http://doi.org/10.21227/y4m0-ka14} \cite{wu2025dataset}.

\begin{figure*}[t]
    \vspace{.8em}
    \centering
    \includegraphics[width=.93\linewidth]{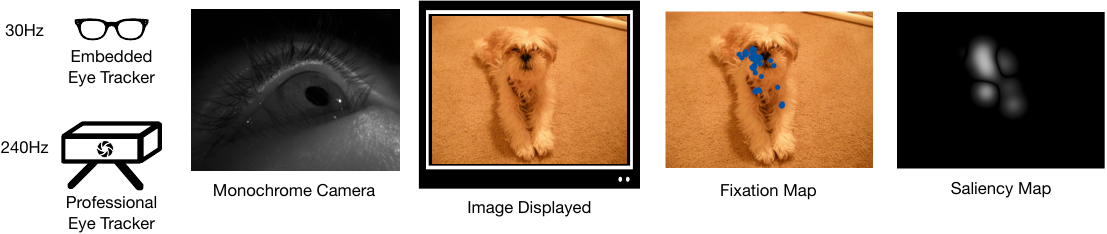}
    \caption{Pipeline of visual fixation prediction. Using recorded videos from the monochrome cameras embedded in the Aria Glasses, we generate both the fixation map and the saliency map.}
    \label{fig:pipeline}
\end{figure*}

\section{MATERIALS AND METHODS}

\subsection{Eye Trackers}\label{sec:tracker}

In the experiments, we utilize Aria Glasses (Meta Platforms, Inc., Menlo Park, CA, USA), which are equipped with two monochrome, global-shutter, inward-facing cameras designed for eye-tracking. These cameras offer a diagonal field of view of 80° and capture images at a resolution of $320 \times 240$ pixels (internally downsampled from $640 \times 480$ pixels) \cite{engel2023project}. The cameras will capture raw eye movement data for subsequent analysis. An additional RGB camera located at the front of the glasses captures the scenes perceived by the subject, with a resolution of $1408 \times 1408$ pixels and 30Hz frame rate.
Although the eye trackers on Aria Glasses (up to 90Hz) have lower accuracy compared to professional ones, which typically operate at frequencies up to 1000Hz and cost up to €20,000, they offer significant advantages in terms of size, cost, and portability. These factors make the device particularly well-suited for real-world applications, where small form factors and affordability are essential considerations. Furthermore, the Aria Glasses provide a more natural, unobtrusive method of eye-tracking, closer to how such technologies might be used in everyday contexts \cite{lv2024aria}. This makes it an ideal tool for investigating eye-tracking applications outside the laboratory setting, where the constraints of high-precision equipment may be less critical.

\subsection{Dataset}
The MIT1003 dataset \cite{judd2009learning} is a widely used benchmark in visual attention research with the 240Hz eye tracker ISCAN ETL-400 (ISCAN Inc., Woburn, MA, USA), containing 1,003 natural images paired with eye-tracking data from 15 participants who viewed the images under free-viewing conditions. This dataset provides valuable insights into human visual fixation patterns and saliency modeling. For our subjective experiments, we uniformly selected 300 images from the MIT1003 dataset to balance the diversity of visual content with the practical constraints of participant availability. This subset allows us to conduct meaningful analyses while ensuring the experiment remains manageable for our limited number of participants. 

\subsection{Protocol}\label{sec:protocol}
In this section, we introduce the detailed protocol of the subjective experiments, which includes the design and the process until we obtain the raw recordings of the eye movements. The post-processing and the evaluation of the visual fixations are presented in Section \ref{sec:experiment}.

We combine this study with a separate investigation focused on animal tracking \cite{mededovic2025eye}, both of which are part of the subjective experiments described in this paper. These two experiments will use the same eye tracker system integrated into the Aria Glasses (Section \ref{sec:tracker}), ensuring consistent data collection methods across the two setups. Both the RGB camera and the eye-tracking camera are set to capture at 30Hz to ensure better synchronization between the two.

The experiments consist of six sessions, each containing 50 images, with each image displayed for 3 seconds. A 1-second interval follows each image, during which a neutral gray screen is shown. Participants will be presented with a variety of images displayed on a 55-inch 4K OLED TV (model OLED55G19LA, LG Electronics, Inc., Seoul, South Korea), ensuring high-quality visual stimuli. Participants are seated approximately 1 meter from the TV screen. According to the calculations in \cite{judd2009learning}, one degree of visual angle corresponds to 27 pixels in our setup (cf. 35 pixels in \cite{judd2009learning}).

Throughout the experiment, participants are instructed to freely view the images with the primary goal of memorizing them, although no formal memory test will be administered afterward. The design of the task encourages relaxed viewing without strict focus on performance following \cite{Judd_2012}. The entire experiment lasts up to one hour, including the introduction, instructions, and breaks between the sessions. 

In total, 10 observers (age range 23-39 years) participated in our study. One result is omitted in this paper due to blurry recordings. In the following, the remaining 9 participants are denoted with S-1 to S-9.

\subsection{Evaluation Metrics}
Evaluating saliency map prediction requires several metrics to capture different aspects of model performance \cite{riche2013saliency,Kummerer2018ECCV,bylinskii2018different}. Commonly used metrics include AUC-Judd \cite{judd2009learning}, Shuffled AUC (sAUC) \cite{tatler_visual_2005}, Normalized Scanpath Saliency (NSS) \cite{peters_components_2005}, Information Gain (IG) \cite{kummerer_information-theoretic_2015}, Correlation Coefficient (CC) \cite{jost_assessing_2005}, Similarity (SIM) \cite{Judd_2012}, and Kullback-Leibler Divergence (KLD), each highlighting unique features of the saliency map's alignment with human gaze.

AUC-Judd \cite{judd2009learning} measures how well the predicted saliency map ranks fixation points by computing the area under the receiver operating characteristic (ROC) curve:
\[
\text{AUC}_{\text{Judd}} = \int_0^1 \text{TPR}(\tau) \, \text{d}(\text{FPR}(\tau)),
\]
where \(\text{TPR}(\tau)\) and \(\text{FPR}(\tau)\) are the true positive and false positive rates at threshold \(\tau\). This metric is widely used for assessing how well saliency models predict fixation locations relative to random locations.

sAUC \cite{tatler_visual_2005} extends AUC-Judd using non-fixation points that are not randomly chosen from the image, but \(N\) shuffled images from other fixations across the dataset, which helps remove the center bias:
\[
\text{sAUC} = \frac{1}{N} \sum_{n=1}^{N} \text{AUC}_n.
\]

The Normalized Scanpath Saliency (NSS) \cite{peters_components_2005} evaluates saliency by comparing predicted saliency values at fixation points to the global mean and standard deviation:

\[
\text{NSS} = \frac{1}{|T|} \sum_{t \in T} \frac{S(t) - \mu_S}{\sigma_S},
\]
where \(\mu_S\) and \(\sigma_S\) are the mean and standard deviation of the saliency map $S$, respectively. $S(t)$ is the predicted saliency value at fixation location $t$ and $T$ is the set of all human fixation points in the image. Higher NSS ($>1.0$) indicates the model assigns significantly higher saliency values to fixation locations than to random locations.

Information Gain (IG) \cite{kummerer_information-theoretic_2015} quantifies how much information the saliency map provides about fixation locations, measuring the reduction in uncertainty between predicted and actual fixation distributions. It also accounts for Uniform (IG-uniform) and Center Bias (IG-centerbias), which are common patterns in human gaze behavior as the baseline ($B(t)$). The IG between the saliency map $S(t)$ and the baseline map \(B(t)\) given the fixation locations $t$ is calculated as:
\[
\text{IG} =  \frac{1}{|T|} \sum_{t \in T}\log_2 \left(\frac{S(t)}{B(t)}\right).
\]
The addition of uniform and center biases allows IG to address the natural biases in human visual attention toward the center and uniform areas of the image. A higher IG indicates better prediction accuracy, but it also corrects for such biases.

Following \cite{riche2013saliency} and \cite{bylinskii2018different}, AUC-Judd, sAUC, NSS, and IG are classified as \textit{location-based} metrics, where saliency values in \( S \) are evaluated based on the locations \( t \) in a binary fixation map. Below, we introduce three \textit{distribution-based} metrics, where two normalized saliency maps, \( S_1 \) and \( S_2 \), are treated as probability distributions. This implies that \( \sum_i S_1 = \sum_i S_2 = 1 \), where \( i \) iterates over all pixels.

The Pearson's Correlation Coefficient (CC) \cite{jost_assessing_2005} measures the linear relationship between the predicted and ground truth saliency maps:
\[
\text{CC} = \frac{\sum_{i} (S_1 - \mu_{S_1}) (S_2 - \mu_{S_2})}{\sigma_{S_1} \sigma_{S_2}} ,
\]
where $\mu_{S_1},\mu_{S_2}$ denote the mean values and $\sigma_{S_1},\sigma_{S_2}$ the standard deviations of $S_1$ and $S_2$, respectively. CC measures the linear relationship between two saliency maps.

SIM \cite{Judd_2012} evaluates spatial overlap between predicted and ground truth saliency maps:
\[
\text{SIM} = \sum_{i} \min(S_1, S_2).
\]
A higher SIM score indicates better alignment.

Finally, Kullback-Leibler Divergence (KLD) measures the difference between the predicted and the ground truth saliency map distributions:
\[
\text{KLD}(S_2 || S_1) = \sum_{i} S_2 \log \frac{S_2}{S_1}.
\]
 Since KLD is not symmetric, meaning \(\text{KLD}(S_2 || S_1) \neq \text{KLD}(S_1 || S_2)\), we report KLD in both directions, denoted as KLD and KLD-rev. A lower KLD suggests better alignment between the predicted and actual saliency distributions.

Together, these metrics provide a thorough evaluation of saliency models by assessing ranking performance (AUC-Judd, sAUC), spatial alignment (NSS, SIM, CC), and information content (IG, KLD), offering a holistic view of how well the model predicts human visual attention. AUC-Judd, sAUC and SIM range within $[0,1]$, CC ranges between $[-1,1]$, while NSS, IG, KLD are unbounded. We adopt the the MATLAB-based implementation from \cite{judd2009learning} and utilize the open-source python package Pysaliency\footnote{\url{https://github.com/matthias-k/pysaliency/}}.

\begin{figure*}[p]
    \centering
    \includegraphics[width=.9\textwidth]{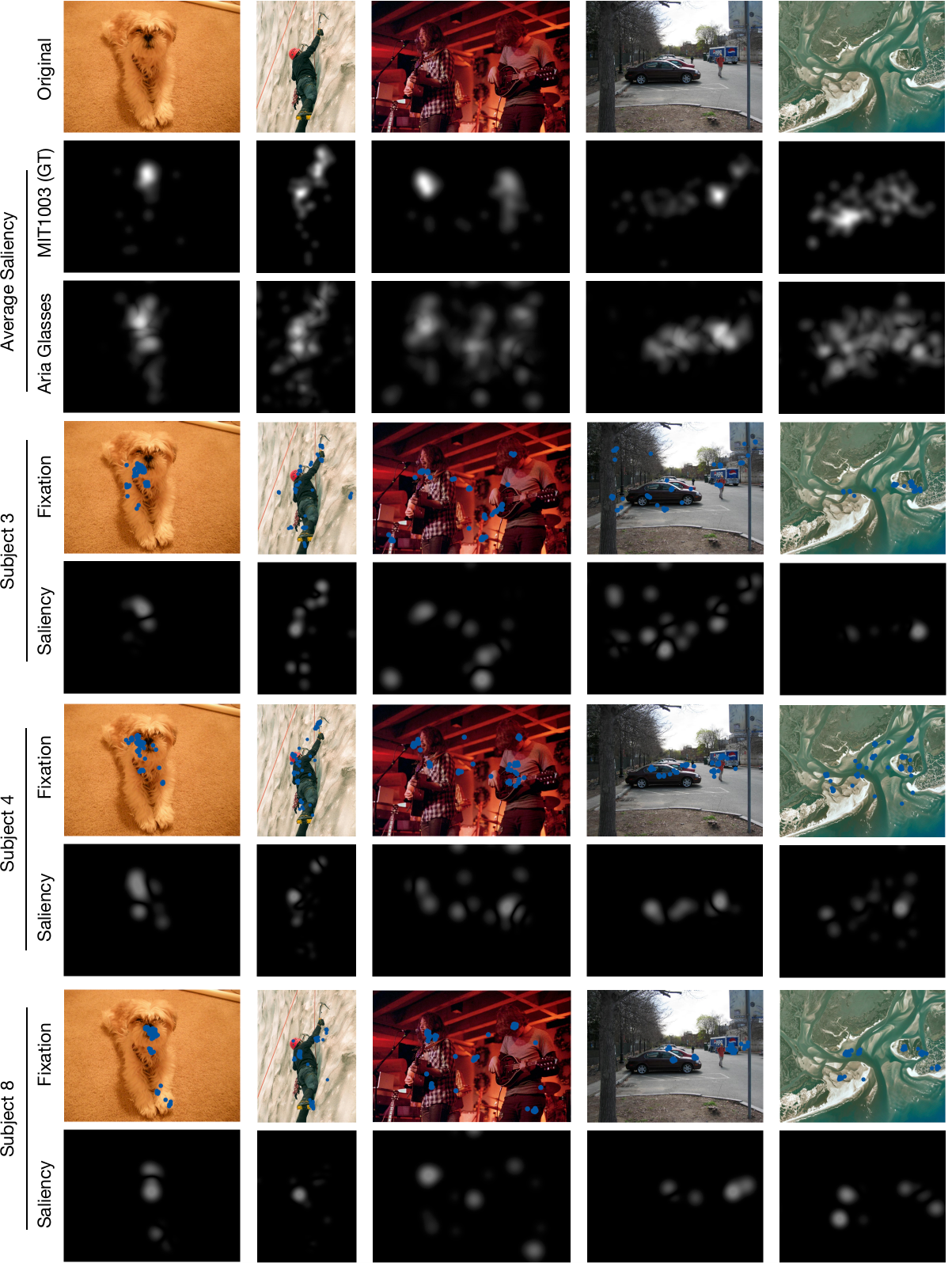}
    \caption{Visualization of fixations and saliency maps. We showcase five images from the MIT1003 dataset, along with the average saliency maps derived from both the MIT1003 ground truth (15 subjects using the ETL 400 ISCAN eye tracker) and our own experiments (9 subjects using the Aria Glasses). For the comparison of visual fixations and corresponding saliency maps, three random subjects are selected from the experimental group.}
    \label{fig:vis}
\end{figure*}

\section{EXPERIMENTS AND RESULTS}\label{sec:experiment}
\subsection{Eye Gaze Prediction}\label{sec:prediction}
The Project Aria Eye Tracking repository\footnote{\url{https://github.com/facebookresearch/projectaria_eyetracking}} provides an open-source (under the Apache 2.0 license) PyTorch-based model designed to estimate eye gaze direction from images captured by the Aria device’s eye-tracking cameras. Trained on a substantial dataset comprising over 1,200 individuals and approximately 2 million frames, this model delivers accurate gaze estimations.
The produced outputs are compatible with the Project Aria Tools Machine Perception Services (MPS) eye gaze output and visualizers, facilitating seamless integration into broader applications. It leverages a ResNet-18 architecture \cite{he2016deep}, consisting of a feature extraction backbone and a prediction head, to predict the eye gaze in the world coordinate given the video frame recorded from two monochrome eye-tracking cameras. 

First, we use the model to predict the gaze coordinates from the eye movement videos. The gaze data, originally in the world coordinate system, is then transformed into the coordinate system of the images from the MIT1003 dataset using a perspective transformation. This transformation maps the quadrilateral borders on the video recorded by the front RGB camera to the in the original image, after applying thresholding and Canny edge detection \cite{canny1986computational}. Therefore, synchronizing the front RGB video with the eye-tracking cameras is crucial, as the homography matrix derived from the RGB videos is applied to the eye gaze coordinates predicted by the eye-tracking cameras. We also tested the Scale-Invariant Feature Transform (SIFT) \cite{lowe1999object}, but it performed worse in comparison.

\begin{figure}[p]
    \vspace{1em}
    \centering
    \begin{minipage}{\linewidth}
        \centering
        \includegraphics[width=0.88\linewidth]{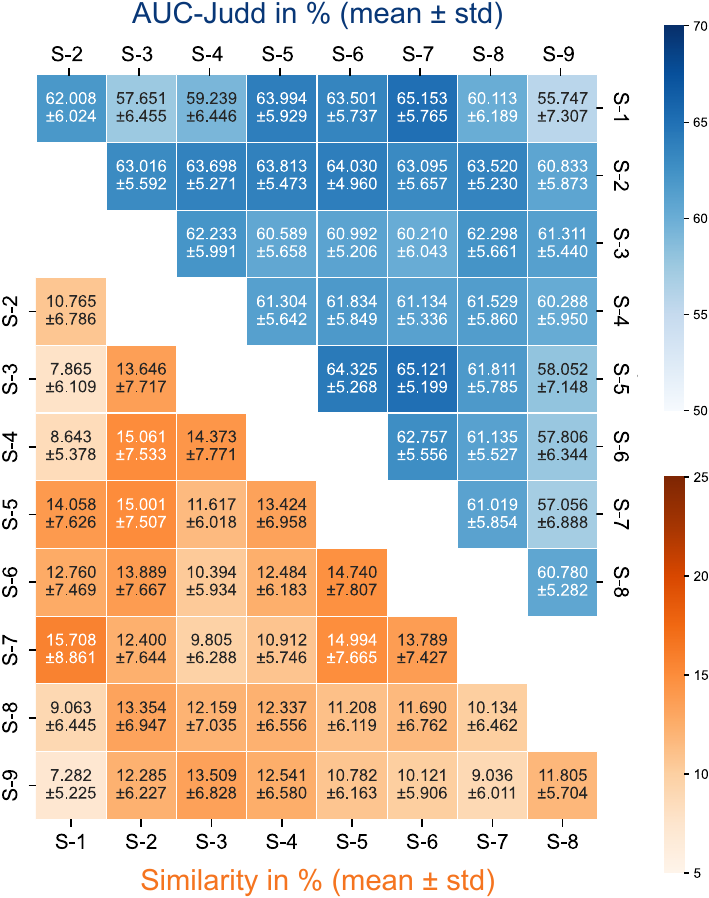}
        \caption{Cross-Subject Consistency. AUC-Judd (upper triangle in blue) and similarity scores (lower triangle in orange) are reported as mean ± standard deviation across individual subject pairs.}
        \label{fig:conf}
    \end{minipage}
    
    \vspace{1.3em} 
    
    \begin{minipage}{\linewidth}
        \centering
        \includegraphics[width=0.92\linewidth]{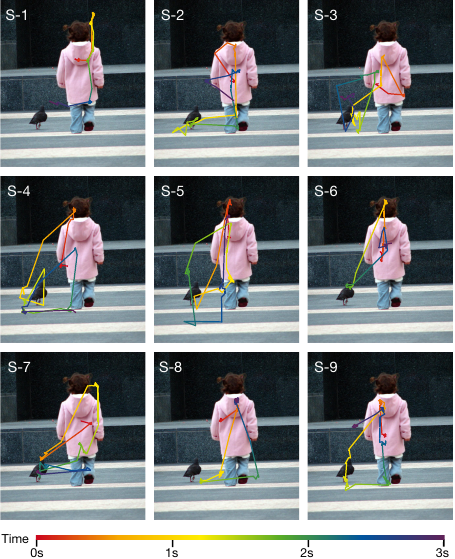}
        \caption{Scanpaths from 9 subjects on an example image. Fixation points are connected sequentially, with a temporal color gradient from red to violet. Best viewed in color and with zoom for details.}
        \label{fig:scanpath}
    \end{minipage}
\end{figure}

To this end, we collected eye gaze data using a 30Hz eye tracker on 300 images from the MIT1003 dataset, with each image viewed for 3 seconds by 9 subjects, resulting in a total of 243,000 eye fixations.

\subsection{Cross-Subject Consistency}
\begin{figure*}[t!]
    \vspace{1em}
    \centering
    \begin{minipage}{0.32\textwidth}
        \centering
        \includegraphics[width=\linewidth]{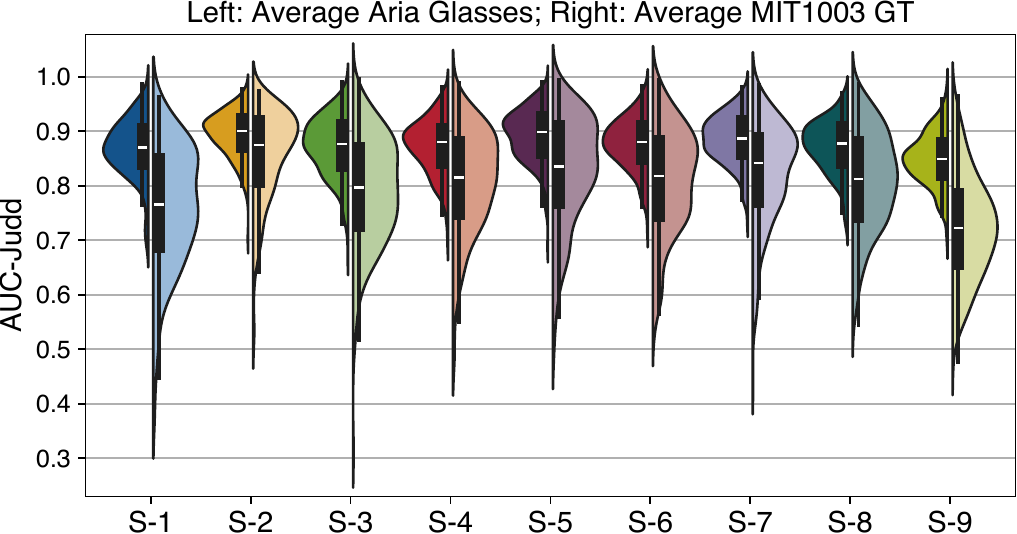}
        \subcaption{}
    \end{minipage}\hfill
    \begin{minipage}{0.32\textwidth}
        \centering
        \includegraphics[width=\linewidth]{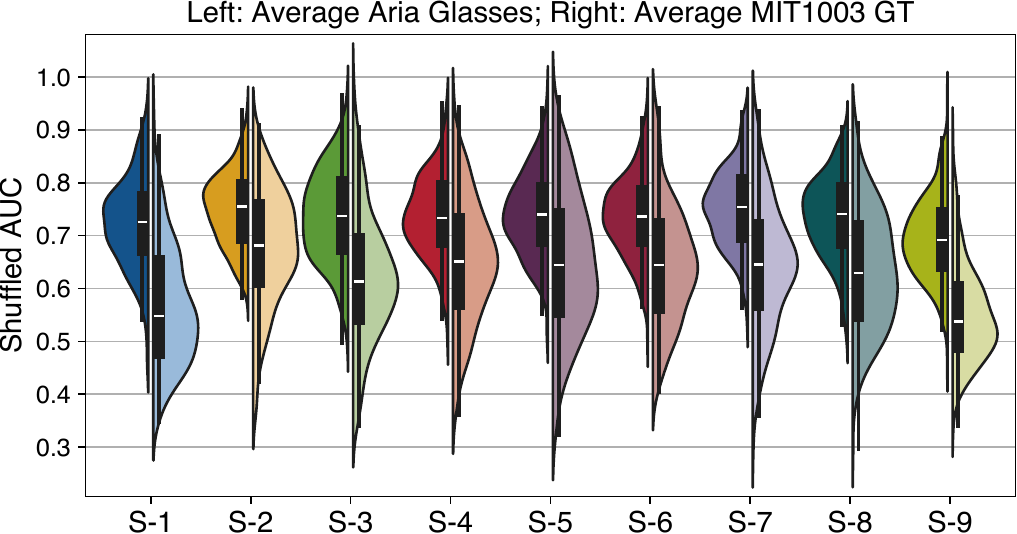}
        \subcaption{}
    \end{minipage}\hfill
    \begin{minipage}{0.32\textwidth}
        \centering
        \includegraphics[width=\linewidth]{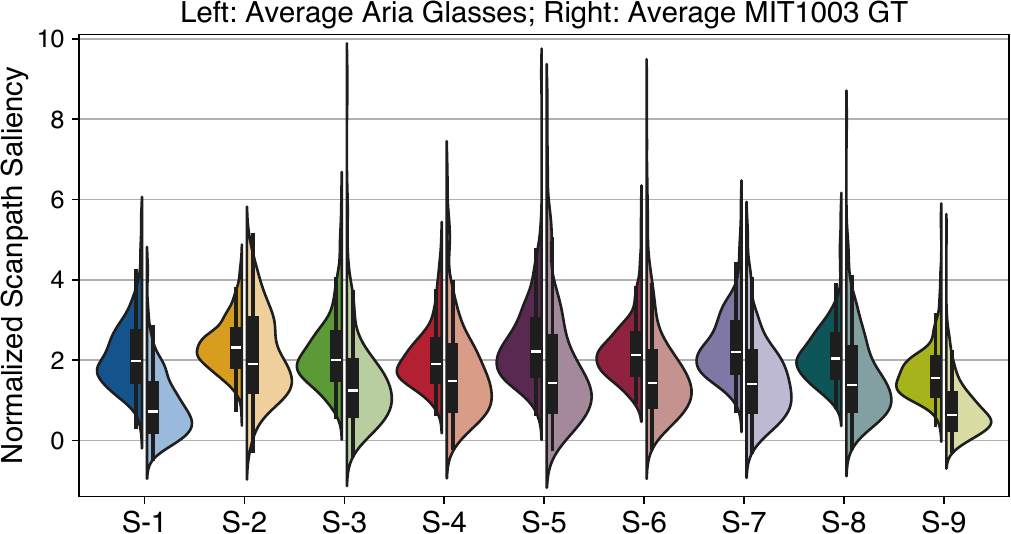}
        \subcaption{}
    \end{minipage}

    \vspace{.8em}

    \begin{minipage}{0.32\textwidth}
        \centering
        \includegraphics[width=\linewidth]{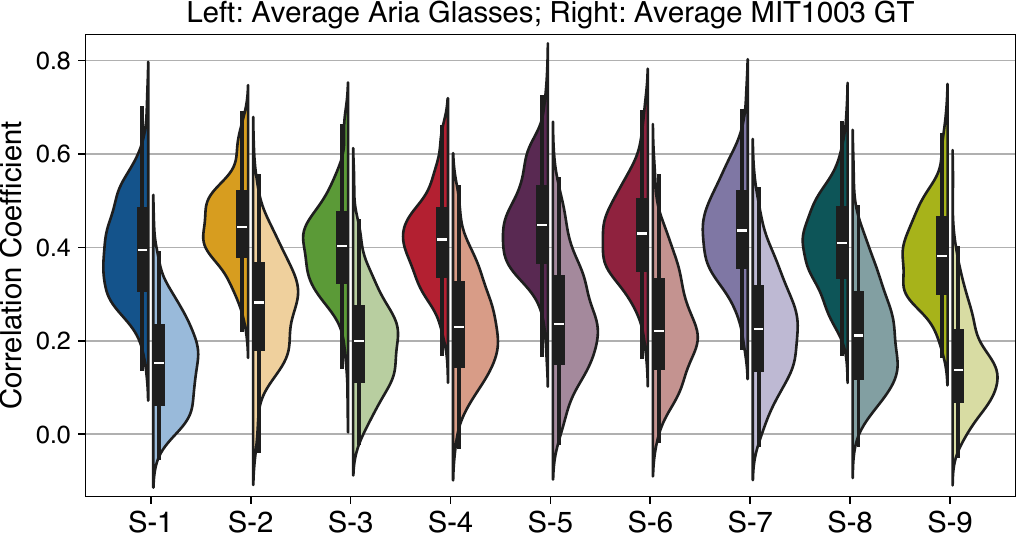}
        \subcaption{}
    \end{minipage}\hfill
    \begin{minipage}{0.32\textwidth}
        \centering
        \includegraphics[width=\linewidth]{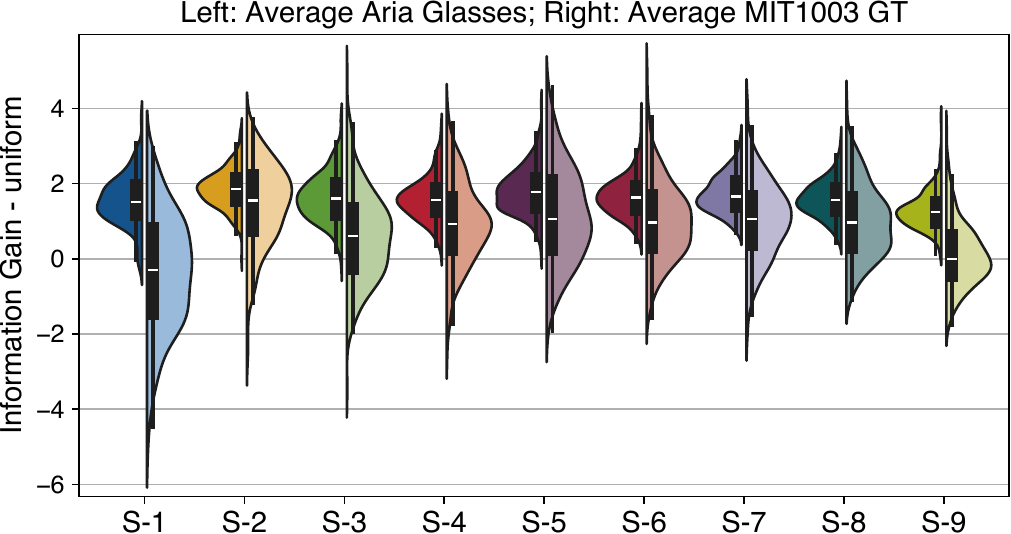}
        \subcaption{}
    \end{minipage}\hfill
    \begin{minipage}{0.32\textwidth}
        \centering
        \includegraphics[width=\linewidth]{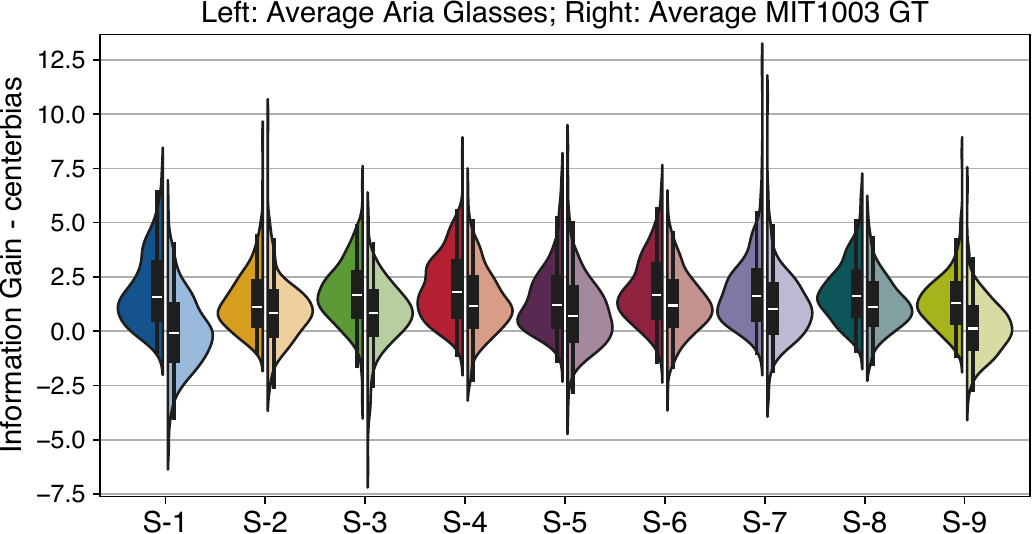}
        \subcaption{}
    \end{minipage}

    \vspace{.8em}

    \begin{minipage}{0.32\textwidth}
        \centering
        \includegraphics[width=\linewidth]{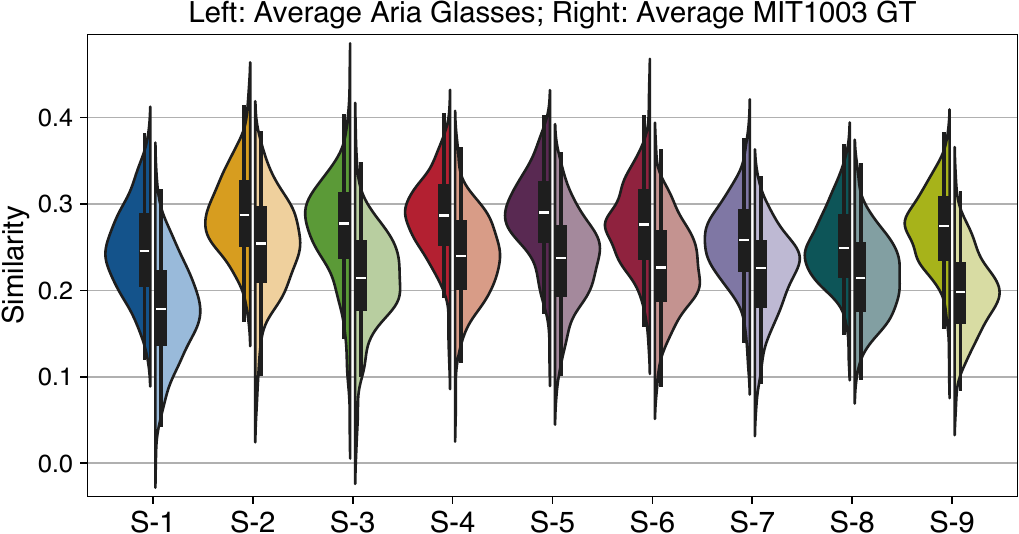}
        \subcaption{}
    \end{minipage}\hfill
    \begin{minipage}{0.32\textwidth}
        \centering
        \includegraphics[width=\linewidth]{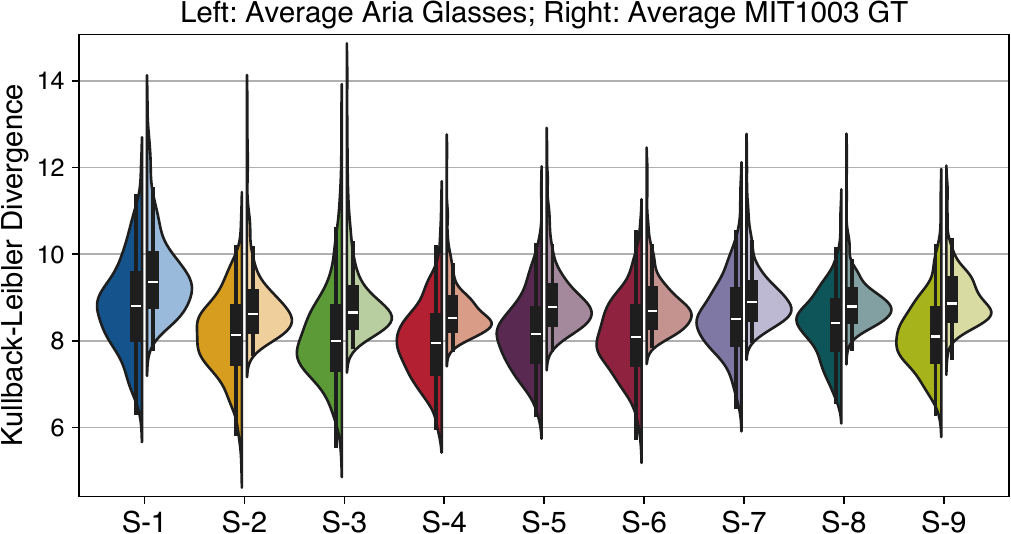}
        \subcaption{}
    \end{minipage}\hfill
    \begin{minipage}{0.32\textwidth}
        \centering
        \includegraphics[width=\linewidth]{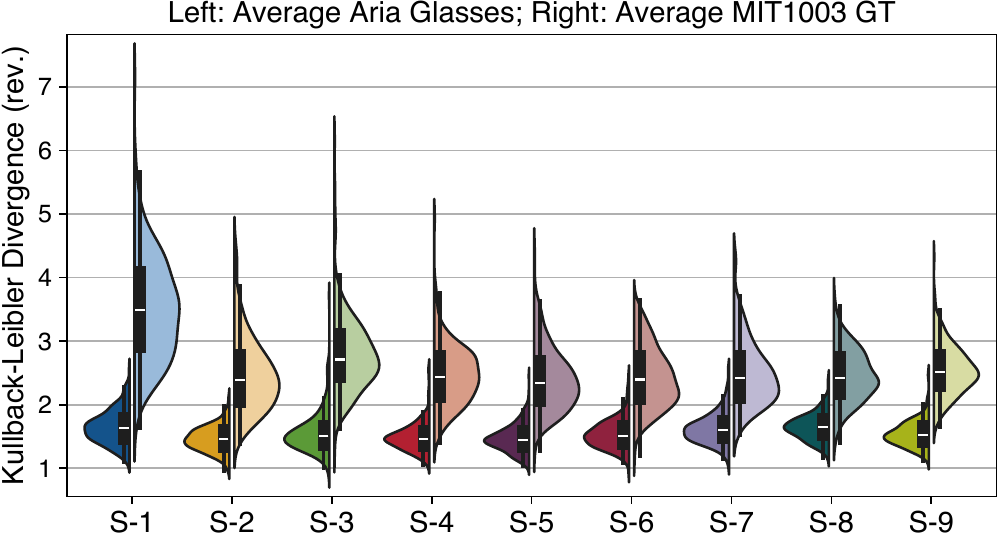}
        \subcaption{}
    \end{minipage}

    \caption{Evaluation metrics from the individual subject to the average using the Aria Glasses (left parts of the violin plots in darker colors) and to the ground truth average using the ETL 400 ISCAN eye tracker from the MIT1003 dataset (right parts of the violin plots in lighter colors).}
    \label{fig:violin}
\end{figure*}

Previous studies, e.g., \cite{judd2009learning}\cite{Judd_2012}\cite{volokitin2016predicting}, have shown that observers tend to fixate on similar locations within an image, particularly when the scene is not overly complex or contains salient objects such as people, animals, or text. In this work, we investigate this phenomenon using a wearable eye-tracking device with a lower recording frequency. To assess cross-subject consistency, we report two complementary metrics: the \textit{location-based} AUC-Judd and the \textit{distribution-based} SIM \cite{riche2013saliency}. As illustrated in Fig.~\ref{fig:conf}, these metrics yield comparable values across different subject pairs, suggesting a degree of consistency among participants. Furthermore, Fig.~\ref{fig:violin} presents all evaluation metrics, comparing individual subjects’ fixation patterns against the average saliency map obtained using the Aria Glasses and the ground truth saliency of the MIT1003 dataset. The consistency in metric values across subjects further supports this observation.

Qualitatively, fixation patterns tend to be more consistent in simple scenes (first three images in Fig.~\ref{fig:vis}) and become less predictable in complex environments (last two images), aligning with prior findings. Additionally, Fig.~\ref{fig:scanpath} visualizes scanpaths from 9 subjects. Notably, the color of the scanpaths on the bird frequently appears green across subjects, indicating strong cross-subject fixation consistency, even in the temporal domain, when presented with a simple stimulus, such as the example image.

\begin{table}[t]
    \vspace{1em}
    \normalsize
    \centering
    \begin{tabular}{ccc}
    \toprule
    Metrics & mean $\pm$ std\;\;\,\; & Baseline \cite{bylinskii2018different}\\\midrule
      AUC-Judd $\uparrow$   & 0.837 $\pm$ 0.051  & 0.80 \\ 
      sAUC $\uparrow$   & 0.681 $\pm$ 0.073  & 0.64 \\
       NSS $\uparrow$ & 1.666 $\pm$ 0.552  & 1.65 \\
      IG-uniform $\uparrow$  & 1.194 $\pm$ 0.556 & -8.49  \\
      IG-centerbias $\uparrow$  & 0.629 $\pm$ 0.982 & - \\
      SIM $\uparrow$  & 0.436 $\pm$ 0.051 & 0.38  \\
        CC $\uparrow$  & 0.496 $\pm$ 0.094  & 0.53 \\
       KLD $\downarrow$ & 4.584 $\pm$ 1.033 & 6.19  \\
       KLD-rev $\downarrow$ & 1.203 $\pm$ 0.213  & - \\
         \bottomrule
    \end{tabular}
    \caption{Cross-Device Consistency. Evaluation metrics are calculated between the average fixation map of 9 subjects using the Aria Glasses and the ground truth average fixation map of 15 subjects using the ETL 400 ISCAN eye tracker in the MIT1003 dataset. The single-observer baseline from \cite{bylinskii2018different} on the complete MIT1003 dataset is included as a reference.}
    \label{tab:res}
\end{table}

\subsection{Cross-Device Consistency}
We first assess cross-device consistency by comparing two average saliency maps: one derived from our experiments with 9 subjects using the Aria Glasses and another from the MIT1003 dataset, which includes 15 subjects recorded with the ETL 400 ISCAN eye tracker.
Table \ref{tab:res} reports the mean and standard deviation of various evaluation metrics using the single-observer baseline on the entire MIT1003 dataset from \cite{bylinskii2018different}, serving as a reference for future studies. Qualitatively, the second and third rows of Fig.~\ref{fig:vis} illustrate cross-device consistency across five example images, showing a high degree of similarity between the saliency maps.

Moreover, Fig.~\ref{fig:violin} compares individual subject saliency maps to the average maps from the Aria Glasses and the MIT1003 dataset. The noticeable difference between the left and right halves of the violin plot suggests that \textit{individual-to-average} consistency is weak, likely due to varying preferences among individuals regarding the complexity of the images in the dataset.


\section{DISCUSSIONS AND CONCLUSIONS}
In our study, we conduct subjective experiments to evaluate cross-subject and cross-device consistency in visual fixation prediction. Our findings suggest that the \textit{average} visual fixations obtained from one device can be transferred to another, provided the stimulus images are relatively simple. However, since \textit{individual-to-average} consistency is not robust, eye-tracking data from a group of subjects is required to generate reliable average fixation and saliency maps. To support further advancements in this field, we have publicly released the eye gaze data from our experiments \cite{wu2025dataset}, enabling researchers to explore new methodologies and validation approaches.

For future work, we plan to assess the performance of foundation models (such as DINO \cite{caron2021emerging} and DINOv2 \cite{oquabdinov2}) in indirectly predicting visual fixations on larger datasets (such as SALICON \cite{jiang2015salicon}) and investigate their potential for improving gaze estimation. Additionally, we will explore the application of fixation information in neuroprosthetic use cases, aiming to enhance visual processing systems for individuals with visual impairments. For more complex scenes, we will first apply scene simplification techniques to generate simpler stimuli.

Looking ahead, individuals with normal vision could store their visual fixation patterns through eye-tracking recordings and use them to train personalized fixation models, which may ultimately assist in the management of visual impairments.

\newpage
\section*{COMPLIANCE WITH ETHICAL STANDARDS}
This study was performed in line with the principles of the Declaration of Helsinki. Ethical approval was granted by the Interfaculty Ethics Committee of RWTH Aachen University (reference number 03/25).

\section*{ACKNOWLEDGMENTS}
We gratefully acknowledge the Project Aria \cite{engel2023project} from Meta Platforms, Inc. for donating the glasses used in this work. We gratefully acknowledge the time, effort, and cooperation of all participants in the subjective experiments.

\bibliographystyle{IEEEtran} 
\bibliography{IEEEabrv,bib}

\begin{thebibliography}{10}
\providecommand{\url}[1]{#1}
\csname url@rmstyle\endcsname
\providecommand{\newblock}{\relax}
\providecommand{\bibinfo}[2]{#2}
\providecommand\BIBentrySTDinterwordspacing{\spaceskip=0pt\relax}
\providecommand\BIBentryALTinterwordstretchfactor{4}
\providecommand\BIBentryALTinterwordspacing{\spaceskip=\fontdimen2\font plus
\BIBentryALTinterwordstretchfactor\fontdimen3\font minus \fontdimen4\font\relax}
\providecommand\BIBforeignlanguage[2]{{%
\expandafter\ifx\csname l@#1\endcsname\relax
\typeout{** WARNING: IEEEtran.bst: No hyphenation pattern has been}%
\typeout{** loaded for the language `#1'. Using the pattern for}%
\typeout{** the default language instead.}%
\else
\language=\csname l@#1\endcsname
\fi
#2}}

\bibitem{yarbus1967eye}
A.~L. Yarbus, \emph{Eye movements and vision}.\hskip 1em plus 0.5em minus 0.4em\relax Springer, 1967.

\bibitem{land1999roles}
M.~Land, N.~Mennie, and J.~Rusted, ``The roles of vision and eye movements in the control of activities of daily living,'' \emph{Perception}, vol.~28, no.~11, pp. 1311--1328, 1999.

\bibitem{li_saliency_2002}
Z.~Li, ``A saliency map in primary visual cortex,'' \emph{Trends in Cognitive Sciences}, vol.~6, no.~1, pp. 9--16, 2002.

\bibitem{torralba2006contextual}
A.~Torralba, A.~Oliva, M.~S. Castelhano, and J.~M. Henderson, ``Contextual guidance of eye movements and attention in real-world scenes: the role of global features in object search.'' \emph{Psychological Review}, vol. 113, no.~4, p. 766, 2006.

\bibitem{golan2020controversial}
T.~Golan, P.~C. Raju, and N.~Kriegeskorte, ``Controversial stimuli: Pitting neural networks against each other as models of human cognition,'' \emph{Proceedings of the National Academy of Sciences (PNAS)}, vol. 117, no.~47, pp. 29\,330--29\,337, 2020.

\bibitem{ma2024eyegaze}
C.~Ma, H.~Jiang, W.~Chen, Y.~Li, Z.~Wu, X.~Yu, Z.~Liu, L.~Guo, D.~Zhu, T.~Zhang, D.~Shen, T.~Liu, and X.~Li, ``Eye-gaze guided multi-modal alignment for medical representation learning,'' in \emph{The Thirty-eighth Annual Conference on Neural Information Processing Systems}, 2024.

\bibitem{jiang2017learning}
M.~Jiang and Q.~Zhao, ``Learning visual attention to identify people with autism spectrum disorder,'' in \emph{Proceedings of the IEEE International Conference on Computer Vision (ICCV)}, 2017, pp. 3267--3276.

\bibitem{koch2024eye}
N.~A. Koch, P.~Voss, J.~M. Cisneros-Franco, A.~Drouin-Picaro, F.~Tounkara, S.~Ducharme, D.~Guitton, and {\'E}.~de~Villers-Sidani, ``Eye movement function captured via an electronic tablet informs on cognition and disease severity in parkinson’s disease,'' \emph{Scientific Reports}, vol.~14, no.~1, p. 9082, 2024.

\bibitem{urano2021visual}
Y.~Urano, A.~Kurosu, G.~Henselman-Petrusek, and A.~Todorov, ``Visual hierarchy relates to impressions of good design,'' in \emph{ACM CHI ’21 Workshop on Eye Movements as an Interface to Cognitive State}, 2021.

\bibitem{wu_fixational_2024}
E.~G. Wu, N.~Brackbill, C.~Rhoades, A.~Kling, A.~R. Gogliettino, N.~P. Shah, A.~Sher, A.~M. Litke, E.~P. Simoncelli, and E.~J. Chichilnisky, ``Fixational eye movements enhance the precision of visual information transmitted by the primate retina,'' \emph{Nature Communications}, vol.~15, no.~1, p. 7964, 2024.

\bibitem{wu2025visual}
Y.~Wu, D.~D.~T. Nguyen, H.~Konermann, R.~Yilmaz, P.~Walter, and J.~Stegmaier, ``Visual fixation-based retinal prosthetic simulation,'' in \emph{IEEE International Symposium on Biomedical Imaging (ISBI)}, 2025, pp. 1--5.

\bibitem{wu2023deep}
Y.~Wu, I.~Kareti{\'c}, J.~Stegmaier, P.~Walter, and D.~Merhof, ``A deep learning-based in silico framework for optimization on retinal prosthetic stimulation,'' in \emph{45th Annual International Conference of the IEEE Engineering in Medicine \& Biology Society (EMBC)}, 2023, pp. 1--4.

\bibitem{wu2024optimizing}
Y.~Wu, J.~Wittmann, P.~Walter, and J.~Stegmaier, ``Optimizing retinal prosthetic stimuli with conditional invertible neural networks,'' \emph{arXiv preprint arXiv:2403.04884}, 2024.

\bibitem{DeepGazeI}
M.~Kümmerer, L.~Theis, and M.~Bethge, ``{Deep Gaze I}: Boosting saliency prediction with feature maps trained on {ImageNet},'' in \emph{ICLR Workshop Track}, 2015.

\bibitem{DeepGazeII}
M.~Kümmerer, T.~S.~A. Wallis, L.~A. Gatys, and M.~Bethge, ``Understanding low- and high-level contributions to fixation prediction,'' in \emph{Proceedings of the IEEE Conference on Computer Vision and Pattern Recognition (CVPR)}, 2017, pp. 4789--4798.

\bibitem{DeepGazeIIE}
A.~Linardos, M.~Kümmerer, O.~Press, and M.~Bethge, ``Calibrated prediction in and out-of-domain for state-of-the-art saliency modeling,'' \emph{arXiv preprint arXiv:2105.12441}, 2021.

\bibitem{DeepGazeIII}
M.~Kümmerer, M.~Bethge, and T.~S.~A. Wallis, ``{DeepGaze III}: Modeling free-viewing human scanpaths with deep learning,'' \emph{Journal of Vision}, 2022.

\bibitem{NIPS2017_3f5ee243}
A.~Vaswani, N.~Shazeer, N.~Parmar, J.~Uszkoreit, L.~Jones, A.~N. Gomez, L.~u. Kaiser, and I.~Polosukhin, ``Attention is all you need,'' in \emph{Advances in Neural Information Processing Systems}, vol.~30, 2017.

\bibitem{caron2021emerging}
M.~Caron, H.~Touvron, I.~Misra, H.~J{\'e}gou, J.~Mairal, P.~Bojanowski, and A.~Joulin, ``Emerging properties in self-supervised vision transformers,'' in \emph{Proceedings of the IEEE/CVF International Conference on Computer Vision (ICCV)}, 2021, pp. 9650--9660.

\bibitem{oquabdinov2}
M.~Oquab, T.~Darcet, T.~Moutakanni, H.~V. Vo, M.~Szafraniec, V.~Khalidov, P.~Fernandez, D.~HAZIZA, F.~Massa, A.~El-Nouby, \emph{et~al.}, ``{DINOv2}: Learning robust visual features without supervision,'' \emph{Transactions on Machine Learning Research}, 2024.

\bibitem{wu2025dataset}
Y.~Wu, H.~Konermann, E.~Mededovic, P.~Walter, and J.~Stegmaier, ``Dataset for cross-device and cross-subject consistency evaluation in visual fixation prediction,'' in \emph{IEEE Dataport}, 2025.

\bibitem{engel2023project}
J.~Engel, K.~Somasundaram, M.~Goesele, A.~Sun, A.~Gamino, A.~Turner, A.~Talattof, A.~Yuan, B.~Souti, B.~Meredith, \emph{et~al.}, ``Project {A}ria: A new tool for egocentric multi-modal {AI} research,'' \emph{arXiv preprint arXiv:2308.13561}, 2023.

\bibitem{lv2024aria}
Z.~Lv, N.~Charron, P.~Moulon, A.~Gamino, C.~Peng, C.~Sweeney, E.~Miller, H.~Tang, J.~Meissner, J.~Dong, \emph{et~al.}, ``Aria everyday activities dataset,'' \emph{arXiv preprint arXiv:2402.13349}, 2024.

\bibitem{judd2009learning}
T.~Judd, K.~Ehinger, F.~Durand, and A.~Torralba, ``Learning to predict where humans look,'' in \emph{IEEE 12th International Conference on Computer Vision (ICCV)}, 2009, pp. 2106--2113.

\bibitem{mededovic2025eye}
E.~Mededovic, Y.~Wu, H.~Konermann, M.~Kopaczka, M.~Schulz, R.~Tolba, and J.~Stegmaier, ``Eye on the target: eye tracking meets rodent tracking,'' 2025, under review.

\bibitem{Judd_2012}
T.~Judd, F.~Durand, and A.~Torralba, ``A benchmark of computational models of saliency to predict human fixations,'' in \emph{MIT Technical Report}, 2012.

\bibitem{riche2013saliency}
N.~Riche, M.~Duvinage, M.~Mancas, B.~Gosselin, and T.~Dutoit, ``Saliency and human fixations: State-of-the-art and study of comparison metrics,'' in \emph{Proceedings of the IEEE International Conference on Computer Vision (ICCV)}, 2013, pp. 1153--1160.

\bibitem{Kummerer2018ECCV}
M.~Kümmerer, T.~S.~A. Wallis, and M.~Bethge, ``Saliency benchmarking made easy: Separating models, maps and metrics,'' in \emph{Proceedings of the European Conference on Computer Vision (ECCV)}, 2018, pp. 702--717.

\bibitem{bylinskii2018different}
Z.~Bylinskii, T.~Judd, A.~Oliva, A.~Torralba, and F.~Durand, ``What do different evaluation metrics tell us about saliency models?'' \emph{IEEE Transactions on Pattern Analysis and Machine Intelligence}, vol.~41, no.~3, pp. 740--757, 2018.

\bibitem{tatler_visual_2005}
B.~W. Tatler, R.~J. Baddeley, and I.~D. Gilchrist, ``Visual correlates of fixation selection: Effects of scale and time,'' \emph{Vision Research}, vol.~45, no.~5, pp. 643--659, 2005.

\bibitem{peters_components_2005}
R.~J. Peters, A.~Iyer, L.~Itti, and C.~Koch, ``Components of bottom-up gaze allocation in natural images,'' \emph{Vision Research}, vol.~45, no.~18, pp. 2397--2416, 2005.

\bibitem{kummerer_information-theoretic_2015}
M.~Kümmerer, T.~S.~A. Wallis, and M.~Bethge, ``Information-theoretic model comparison unifies saliency metrics,'' \emph{Proceedings of the National Academy of Sciences (PNAS)}, vol. 112, no.~52, pp. 16\,054--16\,059, 2015.

\bibitem{jost_assessing_2005}
T.~Jost, N.~Ouerhani, R.~Von~Wartburg, R.~M{\"u}ri, and H.~H{\"u}gli, ``Assessing the contribution of color in visual attention,'' \emph{Computer Vision and Image Understanding}, vol. 100, no. 1-2, pp. 107--123, 2005.

\bibitem{he2016deep}
K.~He, X.~Zhang, S.~Ren, and J.~Sun, ``Deep residual learning for image recognition,'' in \emph{Proceedings of the IEEE Conference on Computer Vision and Pattern Recognition (CVPR)}, 2016, pp. 770--778.

\bibitem{canny1986computational}
J.~Canny, ``A computational approach to edge detection,'' \emph{IEEE Transactions on Pattern Analysis and Machine Intelligence}, no.~6, pp. 679--698, 1986.

\bibitem{lowe1999object}
D.~G. Lowe, ``Object recognition from local scale-invariant features,'' in \emph{Proceedings of the seventh IEEE International Conference on Computer Vision (ICCV)}, vol.~2, 1999, pp. 1150--1157.

\bibitem{volokitin2016predicting}
A.~Volokitin, M.~Gygli, and X.~Boix, ``Predicting when saliency maps are accurate and eye fixations consistent,'' in \emph{Proceedings of the IEEE Conference on Computer Vision and Pattern Recognition (CVPR)}, 2016, pp. 544--552.

\bibitem{jiang2015salicon}
M.~Jiang, S.~Huang, J.~Duan, and Q.~Zhao, ``Salicon: Saliency in context,'' in \emph{Proceedings of the IEEE Conference on Computer Vision and Pattern Recognition (CVPR)}, 2015, pp. 1072--1080.

\end{thebibliography}

\end{document}